\documentclass[twocolumn,showpacs,aps,pra,floatfix,reprint]{revtex4-1}

\usepackage{amsfonts}
\usepackage{amsmath}
\usepackage{amsfonts}
\usepackage{amssymb}

\usepackage{graphicx}
\usepackage{bm}

\newcommand{\mc}[1]{\ensuremath{\mathcal{#1}}}
\newcommand{\ww}[1]{\ensuremath{\stackrel{\xi_{#1}}{\longrightarrow}}}
\newcommand{\wwc}[1]{\ensuremath{\stackrel{\xi_{#1}^*}{\longrightarrow}}}

\allowdisplaybreaks[3]

\begin{document}
\title{Pathway interference in a loop array of three coupled microresonators}

\author{Sandra Isabelle \surname{Schmid}}

\author{Keyu \surname{Xia}(од©исН)}

\author{J\"org \surname{Evers}}
\affiliation{Max-Planck-Institut f\"ur Kernphysik, Saupfercheckweg 1, D-69117
Heidelberg, Germany} 

\date{\today}

\begin{abstract}
A system of three coupled toroidal microresonators arranged in a loop configuration is studied. This setup allows light entering the resonator setup from a tapered fiber to evolve along a variety of different pathways before leaving again through the fiber. In particular, the loop configuration of the resonators allows for an evolution which we term roundtrip process, in which the light evolves from one resonator sequentially through all others back to the initial one. This process renders the optical properties of the system sensitive to the phases of all coupling and scattering constants in the system. We analyze the transmission and reflection spectra, and interpret them in terms of interference between the various possible evolution pathways through the resonator system. In particular, we focus on the phase dependence of the optical properties. Finally, we discuss possible applications for this phase sensitivity induced by the roundtrip process, such as the measurement of the position of a nanoparticle close to one of the resonators, and the measurement of changes in the refractive index between two resonators. Our analytical results for the applications are supported by proof-of-principle calculations based on finite-difference-time-domain solution of Maxwell's equations in two dimensions on a grid. 
\end{abstract}

\pacs{42.60.Da,42.82.Et,42.25.Hz,42.50.Ct}


\maketitle



\section{Introduction}
In recent years, optical microresonators have received considerable attention, since they offer a wide range of applications such as strong-coupling cavity quantum electrodynamics, the modification of spontaneous emission, optical communication, or as sources of light~\cite{N5,review}. A particular promising example combining several of these ideas is the goal of establishing quantum networks~\cite{zoller, quantint}. By now, a large variety of implementations has been achieved~\cite{N5,PhysRevLett.95.067401,painter,PhysRevLett.103.053901,APL85,strong,woggon,park}. Naturally, also the extension to more than one cavity has been suggested, for example, as chains of coupled ring or disc resonators~\cite{N1,N2}, of defects in photonic crystal hosts~\cite{N3,N4}, of coupled resonator spheres~\cite{spheres, cleopop2008, incpro}, or of coupled square resonators on a grid~\cite{Hammer}.

A particular variant of coupled cavities involves two-dimensional arrays of microcavities, which can be used, e.g., to form photonic molecules~\cite{boriskina1, boriskina2}, or optical filters~\cite{painter1,Popovic1, cleopop2008,Harald1}. 
Light entering such an array can take a number of different pathways inside the cavities before leaving the coupled system, and the interference between these different pathways determines the optical properties of the resonator system. 
This in a certain sense can be seen in analogy to an atom with multiple energy levels connected by several driving laser fields. Also the atom can evolve via different pathways. But a particularly interesting case arises if the laser fields are applied to the atom in a so-called closed-loop configuration~\cite{Buckle,PhysRevA.59.2302,PhysRevA.60.4996,PhysRevLett.84.5308,kajari-schroder:013816,PhysRevLett.93.223601,PhysRevLett.93.190502,shpaisman:043812,evers,KoMa1990,KeKoNa1993,MoFrOp2002,PhysRevA.80.063816,PhysRevA.78.051802}. This means that the laser fields are applied such that the atom can evolve in a non-trivial loop pathway from one initial state through the level scheme back to the initial state, e.g., $|1\rangle \to |2\rangle \to |3\rangle \to |1\rangle$  with atomic states $|i\rangle$ ($i\in \{1,2,3\}$). The loop structure induces rich possibilities for interference between the different pathways, and at the same time  renders the optical properties of the atoms sensitive to the phase of the applied driving fields.
\begin{figure}[b]
\centering
\includegraphics[width=0.95\columnwidth]{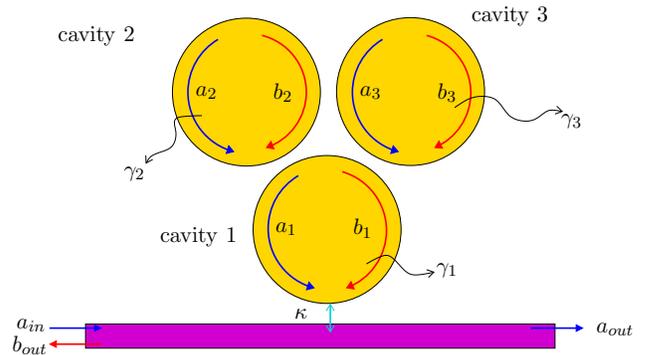}
\caption{\label{skiz}(Color online) The considered setup of three coupled resonators in loop configuration. The 
resonators are probed by a fiber coupled to one of the resonators. Due to the arrangement of the resonators, light can evolve in a roundtrip process, e.g., from cavity 1 via cavities 2 and 3 back to cavity 1 without leaving the resonator array, which leads to rich interference effects. }
\end{figure}
This prompts the question whether similar interference effects and phase-sensitivity could also arise in arrays of microresonators. The couplings between the resonators are mediated via their evanescent fields, and the corresponding coupling constants are in general complex.  Similarly, the scattering inside a given cavity is characterized by a complex scattering constant. Thus it is not surprising that even for simple systems such as two coupled cavities the phase of the couplings can strongly influence the output fluxes. However, there are also cases in which the phases do not influence the final transmission or reflection observed from a resonator, and therefore the coupling constants often are treated as real numbers, neglecting the phase information~\cite{incpro,Popovic1}.

Motivated by this, here, we study array of microresonators in situations in which the phase of the coupling constants are crucial. In particular, we focus on situations in which processes analogous to the closed-loop pathways in atoms occur. For this, we analyze an array of three coupled microresonators probed by a tapered fiber, see Fig.~\ref{skiz}. This setup is the simplest arrangement which allows for a closed-loop roundtrip pathway in analogy to the closed-loop atomic level systems. Due to the arrangement of the resonators, light can evolve in a roundtrip process, e.g., from cavity 1 via cavities 2 and 3 back to cavity 1 without leaving the resonator array. We identify the evolution pathways for the photons entering the resonator array contributing to the transmission and reflection, and determine conditions for the dependence of these optical properties on the phases of the coupling and scattering constants. Based on these results, we analyze the phase dependence for several configurations in detail. Finally, we discuss possible applications for the phase-sensitivity. Our analytical results and interpretations are based on quantum mechanical coupled mode theory. Additionally, we verify mechanism of the proposed applications using numerical finite-difference time-domain solutions of Maxwell's equations in two dimension on a grid.

This article is organized as follows. In Sec.~\ref{sysma}, we describe our model and the observables. In Sec.~\ref{results}, we present our results. We start by analyzing the different pathways light can take through the resonator array, and how phase-sensitivity can arise from these pathways. We then move on to a discussion of transmission and reflection spectra for phase-dependent systems, and of the most important roundtrip process enabled by the loop structure of the resonators. Finally, possible applications are discussed in Sec.~\ref{appl}. Here, we show how to use our setup in order to measure the refractive index of a medium or to determine the position of a nanoparticle.

\section{\label{sysma}Theoretical considerations}

\subsection{Description of the model system}

The system we consider consists of three equal nearby whispering gallery mode microresonators coupled to a tapered glass 
fiber, arranged as shown in Fig.~\ref{skiz}. 
In such a toroidal microresonator a photon can be many times totally reflected at the edges of the cavity and thus move on a polygonal path very similar to a circle. The originally undisturbed resonances of the cavity occur in pairs of clock- and anticlockwise propagating modes. Photons belonging to such a pair $\{a_m,b_m\}$  ($m\in \{1,2,3\}$) differ only by their propagation direction but have the same frequency.
Due to scattering processes, e.g. caused by material imperfections, the modes of such pairs can be scattered into each other. This scattering affects the eigenmodes of the system and thus also changes its eigenenergies. 

As input field we consider a weak probe field of mode $a_{1,in}$, which is coupled into cavity 1. The scattering inside cavity number $n$ we describe by the parameter $h_n$ with $n\in\{1,2,3\}$. The coupling between resonators $n$ and $m$ is described by the coupling constant $\xi_{mn}$ with $m,n\in\{1,2,3\}$. $\kappa$ denotes the coupling strength between the fiber and cavity $1$.
Using these definitions, in a suitable interaction picture  the Hamiltonian of our system reads~\cite{Hamilain}
\begin{equation}
\mc H=\mc H_0+\mc H_L+\mc {H}_{CS}\,,
\end{equation}
with
\begin{subequations}
\begin{align}
{\mc H}_0&=\hbar \sum_{l=1}^3 \Delta_l (a_l^\dagger a_l+b_l^\dagger b_l) \:,\\
 {\mc H}_L&=i\hbar\sqrt{2\kappa} [a_{1,in} a_1^\dagger -a_{1,in}^* a_1 ]\:,\\
 {\mc {H}}_{CS}&=\hbar\sum_{m,n=1}^3 (\xi_{nm} a_n^\dagger b_m+\xi_{nm}^* b_m^\dagger a_n)\:.
\end{align}
\end{subequations}
Here for notational simplicity we defined $h_n=\xi_{nn}$, and used $\xi_{nm}=\xi_{mn}$.
The detunings are defined as $\Delta_l = \omega_l - \omega_{in}$, where $\omega_l$ is the resonance frequency of resonator $l$ and $\omega_{in}$ is the frequency of the probing light. In the following, we assume equal resonance frequencies of the resonators $\Delta_1 = \Delta_2 =\Delta_3=\Delta$ in our calculations.
We write the complex couplings constants as $\xi_{mn}=|\xi_{mn}| e^{i\phi_{mn}}$. For our calculations we assume the critical coupling condition $\kappa=\sqrt{\xi_{11}^2+(\gamma_{1}/2)^2}$ for the coupling between the fiber and cavity 1  to be fulfilled~\cite{turnstile}.
Since the modes $\{a_m,b_m\}$ are assumed to have the same frequency it is reasonable to assume equal internal loss rates $\gamma_1=\gamma_2=\gamma_3=\gamma$ for all modes $\{a_m,b_m\}$. 
Then the total decay of mode $a_1$ $[b_1]$ can be calculated according to $\gamma_{a_1}=2\kappa+\gamma_{1}$ $[\gamma_{b_1}=2\kappa+\gamma_{1}]$, whereas $\gamma_{a_2} = \gamma_{b_2} = \gamma_{a_3} = \gamma_{b_3} =\gamma$.  

From the Heisenberg equation, the time evolutions of the six mode operators $a_m$ and $b_m$ can be obtained as
\begin{subequations}
\label{modedot}
\begin{align}
\dot a_m=&-\left (i\Delta_m + \frac{1}{2}\gamma_{a_m}\right ) a_m-i\sum_{n=1}^3\xi_{mn}b_n \nonumber \\
&+\delta_{m1}\,\sqrt{2\kappa}\,a_{m,in}\:, \\
\dot b_m=&-\left (i\Delta_m + \frac{1}{2}\gamma_{b_m}\right ) b_m-i\sum_{n=1}^3\xi_{nm}^*a_n\:,
\end{align}
\end{subequations}
where $\delta_{ij}$ is the Kronecker Delta function. Since only cavity one couples  to the fiber, in the following we write $a_{1,in}=a_{in}$, $a_{1,out}=a_{out}$ and $b_{1,out}=b_{out}$.

\subsection{Observables}
\label{gencouplings}
In our numerical calculations we investigate the transmission and reflection of light sent into the system via the coupled fiber. In particular, we are interested in the steady state mean values of the output mode operators. We neglect fluctuations  of the photon mode operators and calculate the steady state by setting $\dot a_m = \dot b_m = 0$ in Eq.~\ref{modedot}. For the calculation of the output operators, we use the input-output relation~\cite{MilWals}
\begin{subequations}
\label{ino}
\begin{align}
\langle a_{out}\rangle=&-a_{in}+\sqrt{2\kappa} \langle a_m\rangle\,,\\
\langle b_{out}\rangle=&\sqrt{2\kappa} \langle b_m\rangle\:.
\end{align}
\end{subequations}
Then the transmission and reflection become
\begin{subequations}
\begin{align}
T&=\frac{|\langle a_{out}^\dagger a_{out}\rangle|^2}{|a_{in}|^2} \approx\frac{|\langle a_{out}\rangle|^2}{|a_{in}|^2}\,,\\
R&=\frac{|\langle b_{out}^\dagger b_{out}\rangle|^2}{|a_{in}|^2} \approx\frac{|\langle b_{out}\rangle|^2}{|a_{in}|^2}\:.
\end{align}
\end{subequations}

\section{Results}
\label{results}

\subsection{Light pathway analysis}

\begin{figure}[t]
\centering
\includegraphics[width=0.9\columnwidth]{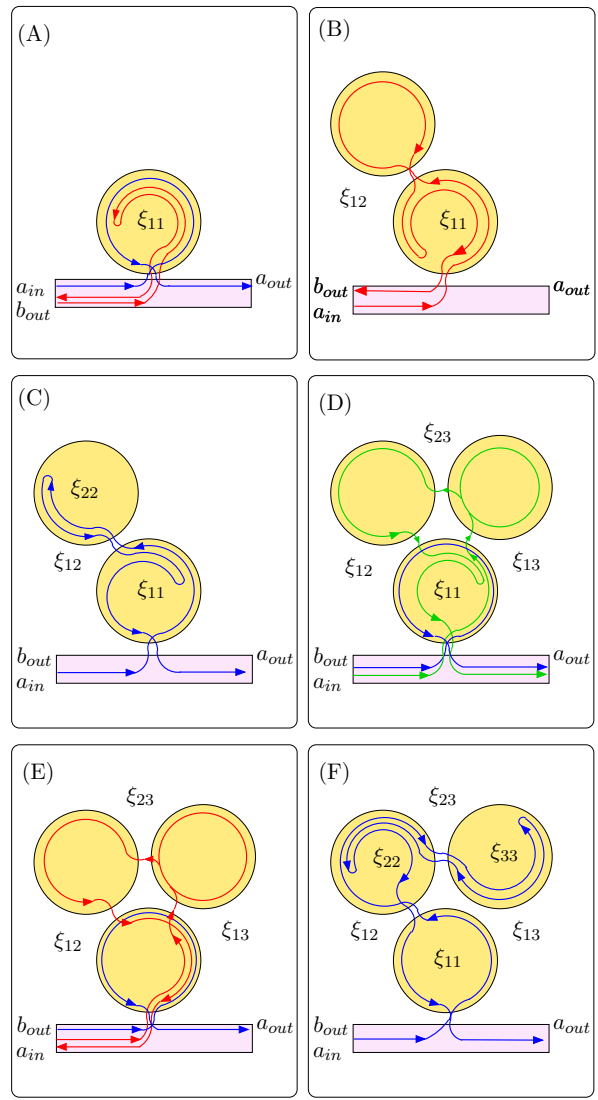}
\caption{\label{path}(Color online) Example pathways of photons passing through the resonator array. In (A), the photon enters only the first cavity, scatters inside cavity 1, and then leaves the cavity back into the fiber. In (B), the photon traverses two cavities, but scatters only in cavity 1. (C) shows a pathway through two cavities with scattering in both the cavities, (D) a path through all cavities with scattering in cavity 1 only, (E) a path through all cavities without scattering, and (F) an evolution through three cavities with scattering in cavities 2 and 3.}

\end{figure}

Light entering our three coupled resonator system can travel along a variety of different pathways before leaving cavity 1 as a transmitted or reflected photon. The different pathways have relative phases, since every time a photon is scattered inside one cavity or moves between two cavities the pathway amplitude is multiplied by the respective complex coupling constant or scattering parameter with phase angle $\phi_{ij}$. The total reflected and transmitted light then arises as the interference of all possible pathways amplitudes. Therefore the phase angles $\phi_{ij}$ can be expected to have a great impact on the resulting output fluxes. In the following, we analyze conditions on the system parameters for obtaining  phase-dependent or phase-independent reflections and transmissions.

We start with a reduced system in which only phase independent pathways are possible and afterwards explain how a phase dependence can arise. For this we set some of the couplings and scattering rates to zero.
If we decouple cavities 2 and 3 from cavity 1 by setting $\xi_{12}=\xi_{13}=0$ as shown in Fig.~\ref{path}(A), no coupling process can take place but only scattering $a_1\stackrel{\xi_{11}^*}{\longrightarrow} b_1$ and vice versa is possible. Although the scattering parameter $\xi_{11}$ has an impact on the light's phase, the overall transmission or reflection are not phase dependent. The reason for this is that all pathways contributing to the transmission include no or an even number of scattering processes, e.g. from $a_1$ to $b_1$ and back. Since the respective scattering parameters are complex conjugates, they have no net influence on the phase of the outcoming flux of each pathways. Similarly, the reflected light constitutes of contributions from pathways with an uneven number of scattering processes. This also means that all interfering contributions have the same phase when leaving the system and thus their interference is constructive independent of the phase of the scattering parameter. 

In the next step, we consider two coupled microcavities by setting $\xi_{12}\neq 0$ whereas the scattering rate $\xi_{22}$ in cavity 2 is assumed to be zero, see Fig.~\ref{path}(B). Taking into account only small numbers of scattering or coupling processes, the following three different pathways are possible:
\begin{subequations}
\label{pathB}
\begin{align}
&a_1\wwc{11}b_1\label{B1}\:,\\
&a_1\wwc{12}b_2\ww{12}a_1\:,\\
&a_1\wwc{12}b_2\ww{12}a_1\wwc{11}b_1\:.
\end{align}
\end{subequations}
Note that we have omitted here pathways differing from the ones in Eqs.~(\ref{pathB}) by a double scattering within cavity 1. In this reduced system, light evolving into cavity 2 always returns to cavity 1 on the same way, and these processes are described by $\xi_{12}$ and $\xi_{12}^*$, such that the phase of the coupling constant $\xi_{12}$ does not affect the output intensities. Similar to the case in Fig.~\ref{path}(A), also the phase of $\xi_{11}$ does not affect the transmission or reflection.

Next, in addition we set $\xi_{22}$ non-zero, see Fig.~\ref{path}(C). In this case, the phases $\phi_{ij}$ do affect the transmission and reflection, since now in addition to the pathways in Eqs.~(\ref{pathB}), two additional leading order pathways are possible:
\begin{subequations}
\label{pathC}
\begin{align}
&a_1\wwc{12}b_2\ww{22}a_2\wwc{12}b_1\label{C1}\:,\\
&a_1\wwc{11}b_1\ww{12}a_2\wwc{22}b_2\ww{12}a_1\:.
\end{align}
\end{subequations}
Thus incoming light can propagate through cavity 1, enter cavity 2, scatter there, and then leave the system in reflection direction. Also, it can pass through cavities 1 and 2 and be scattered in both cavities, leaving the system in transmission direction. These pathways depend on the phase of $\xi_{22}$,  $\xi_{11}$ and $\xi_{12}$, and thus the transmission and reflection become dependent on these phases.

Adding the third cavity, a qualitatively different evolution through the cavity system becomes possible. We first only chose the scattering rate $\xi_{11}$ as nonzero, see Fig.~\ref{path}(D). Now incoming light can evolve on a roundtrip through all three cavities, be scattered in cavity 1 from mode $b_1$ to $a_1$ and afterwards leave the coupled resonator system in transmission direction. Processes like this in which light moves along pathways which describe a loop through all three cavities  will play a special role in our further investigations, as they depend on the phases of all coupling constants between the coupled cavities. In the following we will refer to them as {\it roundtrip processes}. Beside these roundtrip pathways, also non-roundtrip processes are possible, and the transmission and reflection in  leading order are superpositions of the pathways mentioned in Eqs.~(\ref{pathB}) and the following  pathways:
\begin{subequations}
\begin{align}
&a_1\wwc{13}b_3\ww{23}a_2\wwc{12}b_1\label{D1}\;,\\
&a_1\wwc{12}b_2\ww{23}a_3\wwc{13}b_1\label{D2}\;,\\
&a_1\wwc{13}b_3\ww{23}a_2\wwc{12}b_1\ww{11}a_1\label{D3}\;,\\
&a_1\wwc{12}b_2\ww{23}a_3\wwc{13}b_1\ww{11}a_1\;,
\label{pathD}\\
&a_1\wwc{13}b_3\ww{23}a_2\wwc{23}b_3\ww{13}a_1\label{twice1}\;,\\
&a_1\wwc{12}b_2\ww{23}a_3\wwc{23}b_2\ww{12}a_1\label{twice2}\;,
\end{align}
\end{subequations}
plus the pathway in which light enters cavity 1 and leaves the system without any further scattering. The resulting superposition of these amplitudes renders the transmission and reflection dependent on both the phases of the couplings constants $\phi_{ij}$ and the phases of the scattering rate $\phi_{11}$.

It should be noted, however, that without scattering $\xi_{ii}=0$ as in Fig.~\ref{path}(E), also the dependence of the output intensities on the phases of the coupling constants $\phi_{ij}$ disappears. All leading order pathways contributing to the reflection lead the light on a roundtrip pathway [see Eq.~(\ref{D1}) and (\ref{D2})] such that they have the same final phase and thus interfere constructively for all choices of the phase angles $\phi_{ij}$. Similarly, the phase dependence in transmission direction vanishes.

In the most general case with all coupling constants and scattering rates different from zero as indicated in Fig.~\ref{path}(F), many interfering phase dependent pathways become possible, such that a dependence on all phases can be expected. In the following, we will analyze this phase dependence in detail.


\subsection{Transmission and reflection without roundtrip process}

\begin{figure}[t]
\centering
\includegraphics[width=7cm]{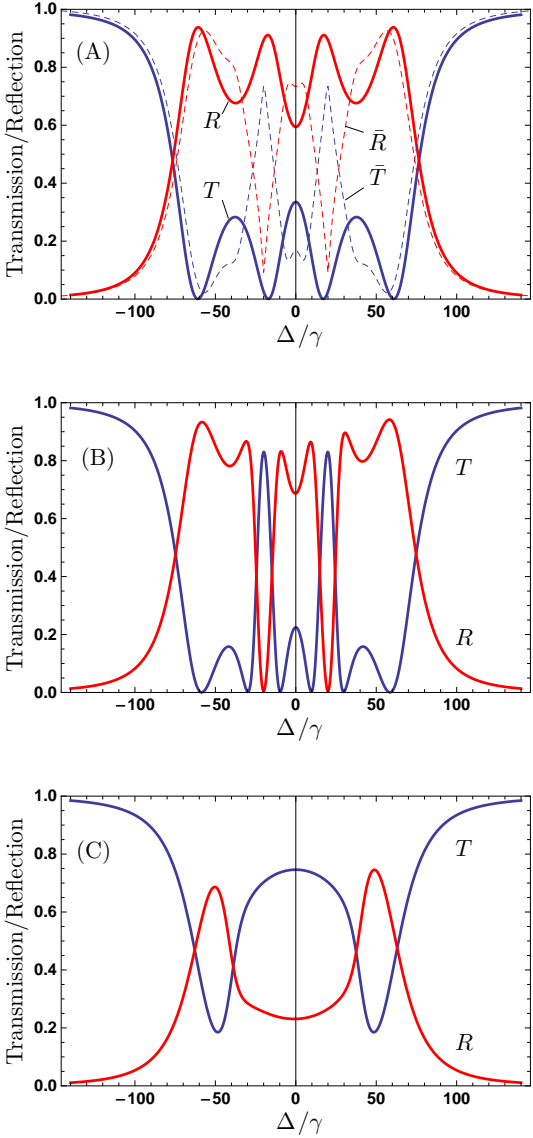}
\caption{\label{outint}(Color online) Transmission $T$ and reflection $R$ in the loop system. The parameters are $\xi_{12}=\xi_{13}=30\gamma$ and $\xi_{23}=0$, such that scattering between cavities $1\leftrightarrow 2$ and $1\leftrightarrow 3$ are possible, but not between 2 and 3. Scattering occurs in all cavities with rates $\xi_{11}=30\gamma$ and $\xi_{22}=\xi_{33}=20\gamma$. In (A) all phases are chosen zero, $\phi_{ij}=0$. The dashed lines show corresponding results for transmission ($\bar T$)  and reflection ($\bar R$) averaged over the phase angles $\phi_{12}$ or $\phi_{22}$. In (B), the phases are chosen as  $\phi_{12}=0.2\pi$, and and all other $\phi_{ij}=0$. In (C), $\phi_{12}=-0.6\pi$ and $\phi_{13}=0.4\pi$, and all other $\phi_{ij}=0$.} 
\end{figure}

We now turn to numerical results for the transmission and reflection in the loop system. In Fig.~\ref{outint} the transmission and reflection are shown for different values of the phase angles $\phi_{ij}$ in dependence of the detuning $\Delta$ of the modes in cavity 1 to the incident light. We choose the scattering rates $\xi_{11}=30\gamma$, $\xi_{22}=20 e^{i\phi_{22}}\gamma$ and $\xi_{33}=20\gamma$. The couplings between the two cavities are $\xi_{12}=30 e^{i\phi_{12}}\gamma$, $\xi_{13}=30 e^{i\phi_{13}}\gamma$ and $\xi_{23}=0$. Since $\xi_{23}=0$, no roundtrip process is possible. For this choice of parameters, the pathways in Eq.~(\ref{pathB}) and (\ref{pathC}) are possible as shown in Fig.~\ref{path}(C), and additionally the analogous ones for cavity 3 instead of cavity 2.

While studying the phase dependence, our variables are the angles $\phi_{12}$, $\phi_{13}$ and $\phi_{22}$. In the solid lines in Fig.~\ref{outint}(A) all $\phi_{ij}=0$, i.e. all coupling and scattering constants are taken as real numbers. In order to explain this result we consider the occupancy and the phase of each light mode inside cavity 1. 
Since coupling light out of the cavity into the glass fiber leads to a phase shift of $\pi$, the outcoupled light of mode $a_1$ interferes constructively with $a_{in}$ when its phase $\phi_{a_1}=\pi$ whereas we obtain destructive interference for $\phi_{a_1}=0$, see also the input-output relations Eq.~(\ref{ino}). As we choose $a_{in}\in\mathbb R$ only the absolute value of $\phi_{a_1}$ is of relevance for analyzing the interference. Therefore we define $\phi_a=|\phi_{a_1}|$. According to the input-output relations  Eq.~(\ref{ino}) we can expect zero transmission $T=0$ if both the phase has a value leading to destructive interference in forward direction, and the amplitude satisfies $2\kappa|a_1|^2=|a_{in}|^2$. If only one of the two conditions is fulfilled, only partial transmission can be expected.
In contrast, for mode b, there is no input field which can interfere with the field leaking out of cavity 1 into reflection direction. Therefore  the reflected intensity is proportional to the intensity of $b_1$ inside cavity 1. 

\begin{figure}[t]
\includegraphics[width=0.9\columnwidth]{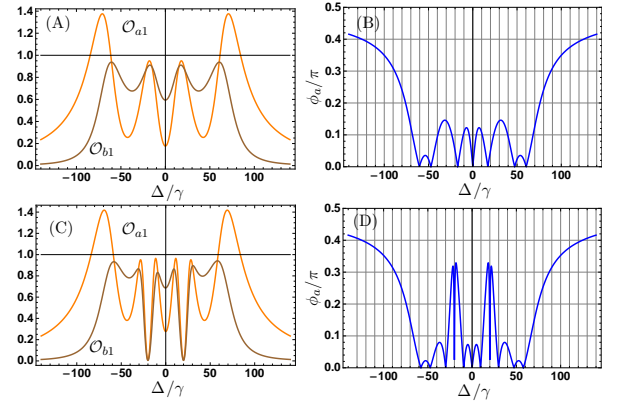}
\caption{\label{intphase}(Color online) Field configuration in resonator 1. (A) and (C): Scaled occupancies $\mathcal{O}_{a_1} = 2\kappa|a_1|^2/|a_{in}|^2$ and $\mathcal{O}_{b_1} = 2\kappa|b_1|^2/|a_{in}|^2$ of the two counter-propagating modes $a_1$  and $b_1$  inside cavity 1. (B) and (D): Phase $\phi_a=|\phi_{a_1}|$ of the counter-clockwise propagating mode inside cavity 1 which is coupled out into the fiber in transmission direction. In (A) and (B), the parameters are chosen as in Fig.~\ref{outint}(A), whereas in (C) and (D) they are as in Fig.~\ref{outint}(B).}
\end{figure}

To verify this interpretation quantitatively, in the upper two subfigures of Fig.~\ref{intphase} we show (A) the scaled photon mode occupancies $\mathcal{O}_{a_1}$ and $\mathcal{O}_{b_1}$ inside cavity 1 proportional to $|a_1|^2$ and $|b_1|^2$, respectively,  and (B) the angle $\phi_a$. Note that the amplitude condition for maximum transmission corresponds to the horizontal line at scaled occupancies equal to one in Fig.~\ref{intphase}(A).
As one can see from Fig~\ref{intphase}(A), the occupancies of both the modes $a_1$ and $b_1$ have maxima around $\Delta=\pm 20\gamma$. By contrast, the transmission in Fig.~\ref{outint}(A) shows minima for these detunings. The reason for this is that $\phi_a(\Delta=\pm 20\gamma)=0$. Thus, the field leaking from $a_1$ into the fiber interferes destructively with the input field $a_{in}$. Almost perfect suppression in forward direction $T\approx 0$ is achieved since for this detuning $2\kappa|a_1|^2\approx |a_{in}|^2$. A similar interpretation holds for $\Delta=\pm 60\gamma$.
For resonant light, i.e. $\Delta=0$, a maximum in $T$ can be observed in Fig.~\ref{outint}(A). This can be traced back to the low occupancy of mode $a_1$ in Fig~\ref{intphase}(A), such that the transmission mainly consists of $a_{in}$. 
Note that for the parameters chosen in Fig.~\ref{outint}(A), both the occupancies and the phases shown in Fig.~\ref{intphase} are symmetric with respect to the detuning. From this it follows that also $T$ and $R$ are symmetric functions of $\Delta$.

In the next example in Fig.~\ref{outint}(B), we in contrast to (A) set $\phi_{12}=0.2\pi$, but keep all other $\phi_{ij}=0$. Compared to the reflection and transmission in (A), we can observe two additional peaks in the transmission and two additional zeros in the reflection signal around $\Delta=\pm 20\gamma$. In order to explain these results, we consider Fig.~\ref{intphase}(C) and (D). 
The minima in $T$ at positions $\Delta=\pm 10\gamma$, $\Delta=\pm 30\gamma$ and $\Delta=\pm 60\gamma$ again arise from destructive interference with $\phi_a=0$ and nearly fulfilled amplitude condition. 
At $\Delta=\pm 20\gamma$ the modes of cavity 1 are nearly unpopulated. Thus $R\approx 0$ and the transmission is governed by the input flux $a_{in}$. Interestingly, cavities 2 and 3 nevertheless contain much higher light intensities, which leads to decoherence via $\gamma$. From $T(\Delta=\pm 20)=0.82\neq 1$ we find that even though  cavity 1 is almost empty, about 20\% of the input light is coupled into the system.
For larger detunings $\Delta$ all results in Fig.~\ref{intphase}(C) and (D) are similar to the graphs in (A) and (B). From this we conclude that in the offresonant case, as expected the phase angles $\phi_{ij}$ of the coupling and scattering parameters have only weak influence on the systems dynamics.
Interestingly, the reflection in Fig.~\ref{outint}(B) is not a symmetric function of the detuning, while the transmission still is symmetric. This difference can arise since different pathways contribute to $T$ and $R$.

In the third example in Fig.~\ref{outint}(C) we chose $\phi_{12}=-0.6\pi$ and  $\phi_{13}=0.4\pi$, but keep all other phases zero. Thus transitions between resonators $1\leftrightarrow 3$ and $1\leftrightarrow 2$ are possible, but not between $2$ and $3$. It can be seen that this change in the phase of the coupling parameters leads to considerable modifications of the transmission and reflection properties. Both transmission and reflection have a simple structure, but are not symmetric with respect to $\Delta$. The interpretation of the peak structure is similar to the two previous cases.

\begin{figure}[t]
\includegraphics[width=0.95\columnwidth]{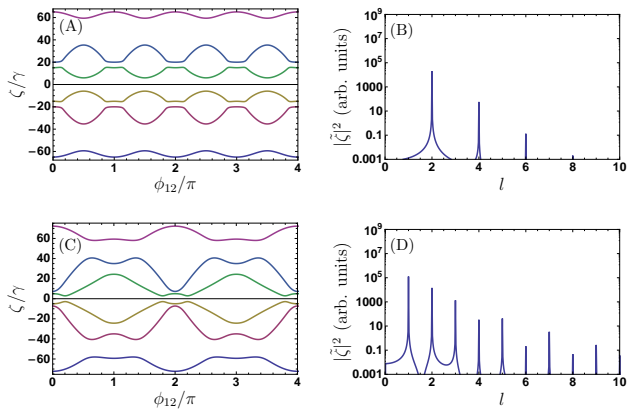}
\caption{\label{EW2}(Color online) Eigenvalue analysis to interpret the phase dependence. The left subfigures (A) and (C) show the imaginary part of the eigenvalues ($\zeta$) of the matrix $\mc M$ describing the system dynamics, which correspond to the energy of the system's dressed states. They are plotted against the phase angle $\phi_{12}$.
The right subfigures (B) and (D) show the power spectrum $|\tilde{\zeta}|^2$ of one eigenvalue as example, revealing the periodicity of the eigenenergies in $\phi_{12}$. The parameters in (A) and (B) are $\xi_{11}=\xi_{13}=30\gamma$, $\xi_{22}=\xi_{33}=20\gamma$, $\xi_{12}=30\gamma e^{i\phi_{12}}$ and $\xi_{23}=0$, thus no roundtrip process is possible. In (C) and (D) $\xi_{23}=15\gamma$, thus the roundtrip process is possible.}
\end{figure}

Analyzing the phase-dependence, we found that the transmission and reflection are $\pi$-periodic in $\phi_{12}$ and $\phi_{13}$ but $2\pi$-periodic in the phases of the scattering parameters $\phi_{ii}$. In the following section, we study and interpret this periodicity of our results in more detail using an eigenvalue analysis.

\subsection{Eigenvalue analysis}

In this section, we analyze the phase-dependence of the transmission and reflection spectra in a more general way. For this, we consider the eigenvalues of the matrix governing our system's dynamics.
The equations of motion for the six mode operators $\{a_i, b_i\}$ can be written as
\begin{subequations}
\begin{align}
\frac{\partial}{\partial t} \vec{C} &=\mc M\cdot \vec{C}\:,\\
\vec{C} &= (a_1,b_1,a_2,b_2,a_3,b_3)^T\,.
\end{align}
\end{subequations}
By diagonalizing $\mc M$, the dressed states of the system can be evaluated. The complex eigenvalues correspond to the complex eigenenergies of these dressed states. The real part of the eigenvalues of $\mc M$ can be interpreted as the decay rates whereas the imaginary parts correspond to the eigenenergies. 
Coupling constants as well as scattering parameters included in $\mc M$ shift the original eigenfrequencies of the resonators. Thus the imaginary parts $\zeta$ of the eigenvalues of matrix $\mc M$ are possible positions for peaks or dips in the reflection and transmission. These eigenvalues thus allow for a study of the dependence of the transmission and reflection spectra on the phase angles $\phi_{ij}$. For example, to study the dependence on $\phi_{12}$, a Fourier transformation of one of the eigenvalues $\zeta$ gives
\begin{align}
\zeta(\phi_{12})=\frac{1}{\sqrt{2\pi}}\int_l \tilde{\zeta}(l) e^{il\phi_{12}} \:\text{d}l\:.
\end{align}
The Fourier coefficients $\tilde{\zeta}(l)$ then determine the periodicity of the eigenvalues in the phase $\phi_{12}$. First, we consider the case where no roundtrip process is possible, i.e. $\xi_{23}=0$. In Fig.~\ref{EW2} we show the six eigenvalues (A) and their power spectrum (B) for the parameters $\xi_{11}=\xi_{13}=30\gamma$, $\xi_{22}=\xi_{33}=20\gamma$, $\xi_{12}=30\gamma\exp[i\phi_{12}]$ and $\xi_{23}=0$. In this case, we observe that $\tilde{\zeta}(l)$ is different from zero only for even numbers of $l$. This means that the eigenvalues are $\pi$ periodic in $\phi_{12}$. If no roundtrip process can take place, light evolving from cavity 1 into cavity 2 or 3 has to move the same way back in order to leave the system and to be detected as output light. Thus it interacts two times with the same coupling constant. This is the reason why the Fourier coefficient belonging to $\phi_{12}$ vanishes for uneven numbers of $l$.
In Fig.~\ref{EW2}(C) and (D) we show the eigenvalues for $\xi_{23}=15\gamma$. All further parameters are the same as before. This setup corresponds to Fig.~\ref{path}(F). In this case, a roundtrip process is possible, and the Fourier coefficients $\tilde{\zeta}(l)$ are different from zero both for even or uneven numbers of $l$. The reason is that light can evolve, e.g., from cavity 1 via cavities 2 and 3 back to 1 such that the phase $\phi_{12}$ influences the path amplitude only  once. Thus in this case, the eigenenergies are $2\pi$ periodic in $\phi_{12}$.

Similar analysis allows to also reveal the phase dependence of the other coupling constants.

\subsection{The roundtrip process}

In this Section we study the roundtrip process and the resulting effects in detail. For this, we make use of the fact that by turning the coupling $\xi_{23}$ on and off it can be controlled whether or not a roundtrip pathway can be taken by entering light. Therefore, in the first step we expand the transmission $T$ in this coupling constant around $\xi_{23}=0$. This expansion will reveal the leading order effects of the roundtrip process for small $\xi_{23}$. Afterwards we present numerical results for $T$ and $R$ for more general parameter sets which allow the roundtrip processes to take place.

\subsubsection{Expansion in orders of the roundtrip process}

\begin{figure}[t]
\includegraphics[width=\columnwidth]{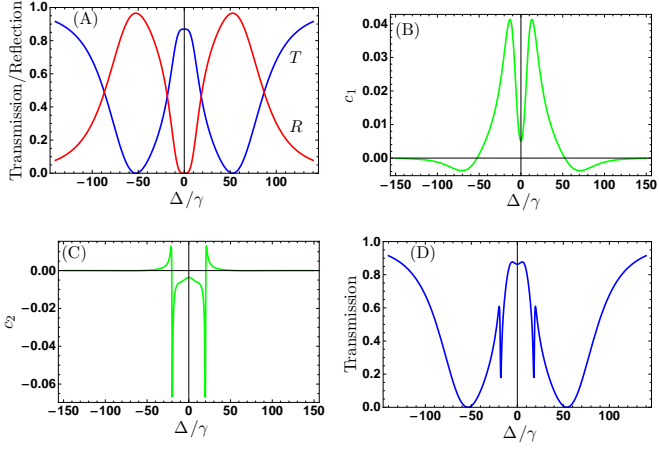}
\caption{\label{taylor}(Color online) Expansion of the transmission in orders of the roundtrip process. For this, the transmission is expanded around 0 in $|\xi_{23}|/\gamma$. The parameters are $\xi_{11}=50\gamma$, $\xi_{22}=20 e^{i\phi_{22}}\gamma$, $\xi_{33}=20\gamma$, $\xi_{23}=10\gamma$, $\xi_{13}=30 e^{i\phi_{13}}\gamma$ and all $\phi_{ij}=0$. (A) zero order contribution ($\xi_{23}=0$), (B) first order contribution $c_1$, (C) second order contribution $c_2$, (D) complete transmission for $\xi_{23}=3\gamma$.}

\end{figure}

To elucidate the impact of the roundtrip process on the transmission spectrum, we performed a Taylor expansion of the transmission amplitude to the second order in $|\xi_{23}|/\gamma$ around $\xi_{23}=0$. This refers to the case where the coupling between cavity 2 and 3 is much weaker than the other couplings $\xi_{ij}$. The expanded transmission reads:
\begin{align}
\frac{|\langle a_{out}\rangle|^2}{|\langle a_{in}\rangle|^2} \approx 
\left |c_0+c_1\frac{|\xi_{23}|}{\gamma}+c_2\frac{|\xi_{23}|^2}{\gamma^2} \right|^2\:.
\label{tay}
\end{align}
Here, $c_n$ are the Taylor expansion coefficients. The  0th order corresponds to the case where no roundtrip process is possible, i.e. $\xi_{23}=0$. The respective result is shown in Fig.~\ref{taylor} (A). The parameters are chosen as $\xi_{11}=50\gamma$, $\xi_{22}=20 e^{i\phi_{22}}\gamma$, $\xi_{33}=20\gamma$, $\xi_{12}=10\gamma$, $\xi_{13}=30 e^{i\phi_{13}}\gamma$ and all $\phi_{ij}=0$. 
In Fig.~\ref{taylor}(B) and (C) we show the first and second order Taylor coefficients $c_1$ and $c_2$ which provide the respective correction terms for the transmission. The first order corresponds to a pathway in which light interacts once with the coupling constant $\xi_{23}$. The most probable process of this kind that ends in $a_1$ is a single roundtrip through all three cavities, see e.g. Eq.~(\ref{D3}) and (\ref{pathD}). The second order contributions correspond to pathways where light moves twice between cavity 2 and 3, including a double roundtrip process. Examples for such pathways are given in Eq.~(\ref{twice1}) and (\ref{twice2}).

We can see from these figures that the first order correction leads to a small dip that overlaps with the broad resonance round $\Delta=0$. By contrast, the second order correction consists of two sharp resonances round $\Delta=20\gamma=\xi_{22}=\xi_{33}$.  Comparing the sum of the curves of Fig.~\ref{taylor} (B) and (C) to the full output for $\xi_{23}=3\gamma$ (D), we find that for the small value of $\xi_{23}$ taken in our example the first two orders in the expansion are sufficient to almost perfectly approximate the full result in (D).

\begin{figure}[t]
\includegraphics[width=\columnwidth]{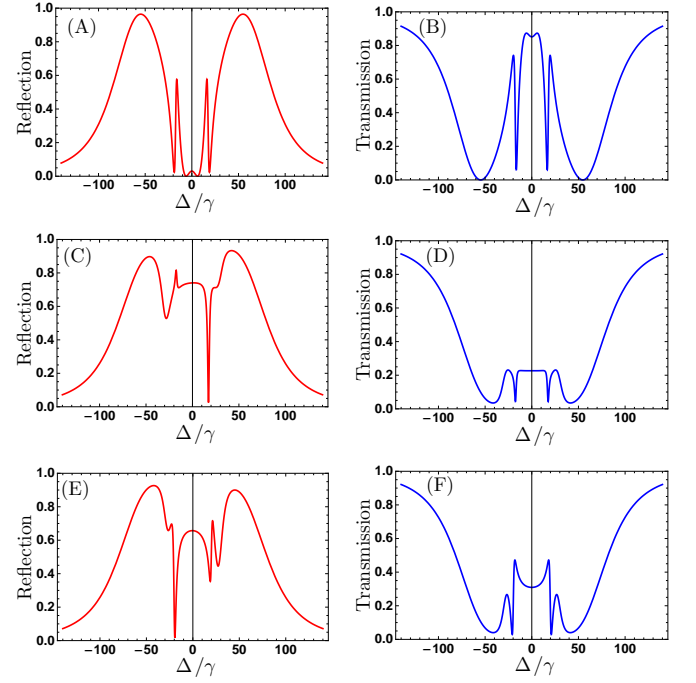}
\caption{\label{loopdep}(Color online) Transmission (blue) and reflection (red) spectra for parameters which allow for the roundtrip process. The parameters are the same as in Fig.~\ref{taylor} except for $\xi_{23}=5\gamma$. In (A) and (B), all phases are zero ($\phi_{ij}=0$). In (C) and (D), $\phi_{22}=1.6\pi$ and $\phi_{13}=0.4\pi$ and all other phases are chosen zero. In (E) and (F), $\phi_{22}=1.6\pi$ and $\phi_{13}=-0.4\pi$ and all other $\phi_{ij}=0$.}
\end{figure}

The structure of $T$ can be explained using the same interpretation techniques as applied for Fig.~\ref{outint} based on the occupancy and phase of the fields inside cavity 1.

\subsubsection{Transmission and reflection with roundtrip process}

\begin{figure}[t]
\includegraphics[width=0.9\columnwidth]{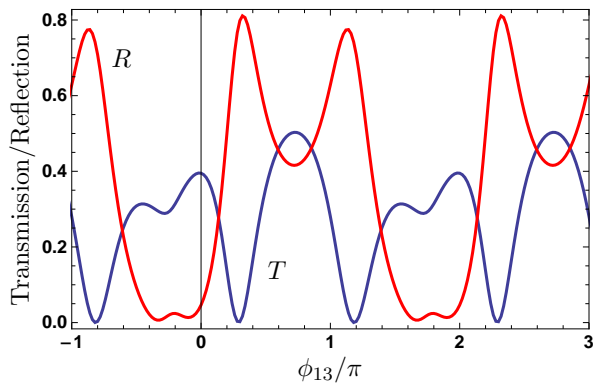}
\caption{\label{phi13}(Color online) Transmission (blue) and reflection (red) spectra for parameters which allow for the roundtrip process. The spectra are plotted against the phase $\phi_{13}$. The other parameters are chosen as in Fig.~\ref{loopdep} with $\Delta=-19.5\gamma$.}

\end{figure}

Next we study the phase dependence of the transmission $T$ and reflection $R$ for a choice of parameters for which a roundtrip process is possible. For this, we choose all coupling constants and all scattering parameters non-zero, as shown in Fig.~\ref{path}(F). All parameters are as in Fig.~\ref{taylor} except for $\xi_{23}=5\gamma$. 
We start by considering the transmission and reflection if all phase angles $\phi_{ij}=0$. The respective results are shown in Fig.~\ref{loopdep}(A) and (B). We can see clearly the two side band dips around $|\Delta|=20\gamma=|\xi_{22}|=|\xi_{33}|$ arising from the second order Taylor correction of Eq.~(\ref{tay}). 

Fig.~\ref{loopdep}(C) and (D) show corresponding results with phase angles changed to $\phi_{22}=1.6\pi$ and $\phi_{13}=0.4\pi$. We observe that the reflection becomes asymmetric and only one sharp dip around $\Delta=20\gamma$ remains. The transmission also changes, but remains symmetric. Upon changing the sign of phase $\phi_{13}$, the sharp dip moves to the opposite side of the spectrum, i.e., to $\Delta\approx -20\gamma$, see subfigures (E) and (F). The transmission still remains symmetric, but is also affected by the phase change of $\phi_{13}$.
In Fig.~\ref{phi13} we show $T$ and $R$ in dependence on the phase angle $\phi_{13}$ for $\Delta=-19.5\gamma$ which is the position of one of the narrow structures. We find that even for small changes of the angle $\phi_{13}$, the transmission and reflection can changes considerably. 

We thus conclude that the possibility of taking a roundtrip pathway in the loop system is the origin of narrow structures in both the transmission and the reflection, and these narrow structures are sensitive to the phases of the coupling constants. This invites applications based on the dependence of these coupling constants on an observable. In the following Section we describe two possible applications based on the sensitivity of the transmission spectrum on the phase angles $\phi_{ij}$.

\section{\label{appl}Applications}

\begin{figure}[t]
 \centering
\includegraphics[width=0.9\linewidth]{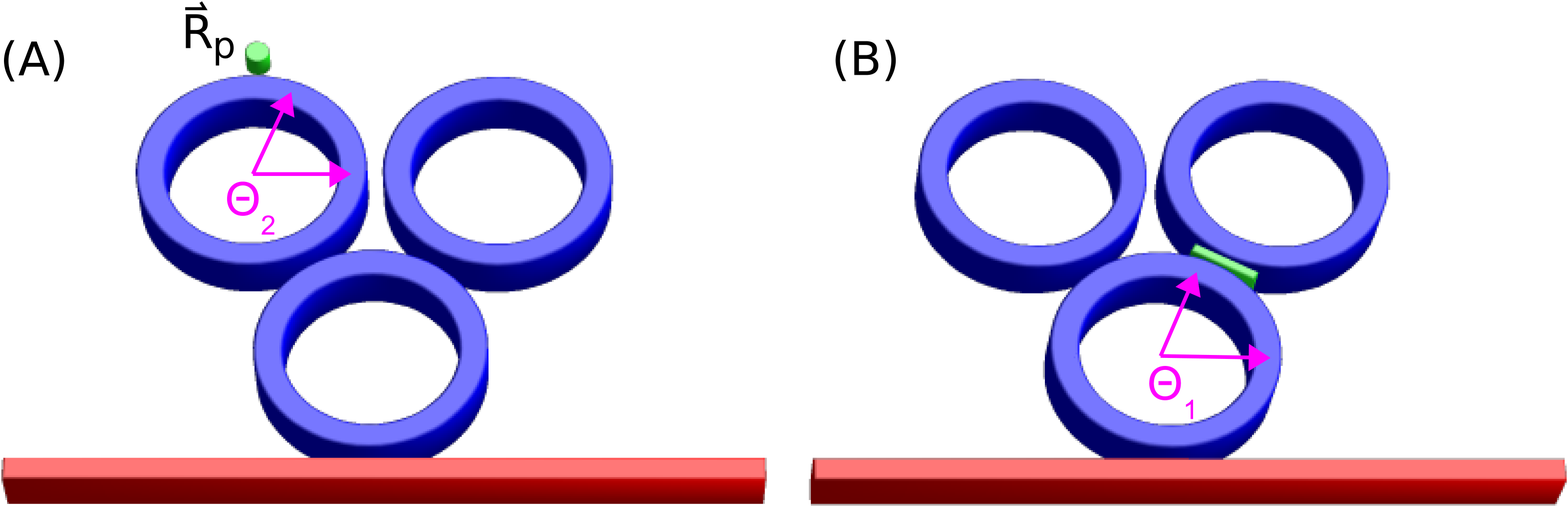}
 \caption{\label{fig:Carton}(Color online) Setups for the two example applications. (A) shows the measurement of the position of a particle at position $\vec{R}_p$ close to resonator 2. (B) shows the measurement of changes in the index of refraction between resonators 2 and 3 induced, e.g., by slabs of different refractive indices.}
\end{figure}

In this Section we discuss two possible applications of our setup. They rely on the dependency of the transmission and reflection on the scattering and coupling parameters. First, we analyze the possibility to detect the position of a small particle such as an atom or a nano object close to one of the resonators as shown in Fig.~\ref{fig:Carton}(A).  Second, we aim at measuring small changes in the refractive index between two cavities. These changes could be induced by an object placed in the free space, or by embedding the interface area in a liquid. The corresponding setup is sketched in Fig.~\ref{fig:Carton}(B).

In the following, we first provide a theoretical background to our calculations, and then discuss numerical results both based on the coupled mode theory and on a numerical integration of Maxwell's equations on a grid.

\subsection{Monitoring the position of a nano particle}

A subwavelength refractive object located very close to one of our cavities as shown in Fig.~\ref{fig:Carton}(A) gives rise to a scattering of the fields propagating inside the cavities. This results in damping and in a coupling between conterpropagating WGM pairs $\{a_n,b_n\}$~\cite{NaturePhoton4p46,PRL99p173603}.
The scattering resulting from a nano particle has been studied in~\cite{PRA75p023814,NaturePhoton4p46}, where it was demonstrated that the size of a nano particle can be determined using high-Q WGM. This raises the question, whether also the particle position can be determined.
To address this question, we note that in the coupled mode theory, the complex electric fields of the modes $a_n$ and $b_n$ in the $n$-th cavity are related and can be written in cylindrical coordinates $\mathbf{R}=(\rho,\theta,z)^T$ as~\cite{PRA75p023814}
\begin{subequations}
\begin{align}
 \mathbf{E}_{b}^0  ( \mathbf{R})&= \left(E_\rho^0(\rho,z),iE_{\theta}^0(\rho,z),E_z^0(\rho,z) \right ) ^T\,e^{im\theta}\\ 
 \mathbf{E}_{a}^0  (\mathbf{R})&= \left(E_\rho^0(\rho,z),-iE_{\theta}^0(\rho,z),E_z^0(\rho,z) \right ) ^T\,e^{-i m\theta}\,.
\end{align}
\end{subequations}
The origin of the used coordinate system is at the center of the WGM cavity.
For high-Q WGMs with small loss rate, $a_n$ and $b_n$ are to a good approximation complex conjugates of each other, such that the three components $E_\rho^0,E_\theta^0,E_z^0$ are real functions~\cite{PRA75p023814}.
We assume that a scattering particle at position $\mathbf{R_p}$ giving rise to a point-like dielectric fluctuation with size much smaller than the wavelength. 
We denote the  dielectric constant of the resonators as $\varepsilon_c$, that of the homogeneous medium surrounding the resonators as  $\varepsilon_s$, and that of the medium with added particle as $\varepsilon_p(\mathbf{R})$. The scattering parameter $\xi_{nn}$ is the proportional to the difference $\delta \varepsilon(\mathbf{R})=\varepsilon_p(\mathbf{R})-\varepsilon_s$ as well as to the intensity of the electric field at the position of the particle, and can be written as~\cite{PRA75p023814,OE13p1515}
\begin{subequations}
\begin{align}
\label{eq:SCT}
 \xi_{nn}& =\frac{\omega_m}{2}\frac{\int_{V_p}(\varepsilon_p(\mathbf{R})-\varepsilon_s)\mathbf{E}_{a}^{0*}(\mathbf{R})
\mathbf{E}_{b}^{0}(\mathbf{R}) d\mathbf{R}}{\int \varepsilon_s |\mathbf{E}_n^{0}(\mathbf{R})|^2 d\mathbf{R}}\\[2ex] 
& \propto \delta \varepsilon(\mathbf{R_p})e^{2im\theta_n(\mathbf{R_p})}|\mathbf{E}^0_n(\mathbf{R_p})|^2 \,.
\label{subb}
\end{align}
\end{subequations}
Here, $m$ is the azimuthal mode number, and in Eq.~(\ref{subb}), for simplicity, we set  $\mathbf{E}^0(\mathbf{R})=\mathbf{E}_{b}^0(\mathbf{R})=\mathbf{E}_{a}^{0*}(\mathbf{R})$.
It can be seen that the scattering resulting from the particle is a complex number. Its phase depends on the position $\theta_n(\mathbf{R_p})$ of the particle, see Fig.~\ref{fig:Carton}(A).  According to Eq.~(\ref{subb}), this angle enters the phase of the coupling constants via $2m\theta_n(\mathbf{R_p})$, such that an increase of the azimuthal mode number leads to higher position sensitivity, but at the cost of a smaller range of uniquely determined positions, since the phase is only determined modulo $2\pi$. Since we found in the previous sections that the transmission and the reflection in our loop setup is sensitive to the phase of the coupling constants, in principle, a position determination becomes possible.

\subsection{Monitoring the dielectric constant of a nano slab}

Next, we turn to the measurement of the dielectric constant of a thin object located in the space between two of the cavities. Alternatively, this slight change could also be induced by the concentration of a fluid between the two resonators, or by varying the temperature~\cite{Science317p783}. 
In general, modifying the dielectric constant in between the two cavities gives rise to a change in the coupling of two cavities and to scattering. However, for a larger sample exceeding the wavelength scale, the scattering can be small such that the change in the coupling constant is dominant. In the following, we assume this condition to be fulfilled and neglect the scattering induced by the slab, and consider a slab in the region $V_{slab}$.
The total coupling can then be separated into three parts: The coupling without slab $\xi_{nm}^{(0)}$, the contribution $\xi_{nm}^{(slab)}$ from the slab with a reference constant $\varepsilon_{slab}^0$ inserted in the gap of two cavities, and a change $\delta\xi_{nm}^{(slab)}$ as the dielectric constant of the slab varies according to $\delta\varepsilon_{slab} = \varepsilon_{slab}-\varepsilon_{slab}^0$.
Thus the total coupling is given by
\begin{subequations}
\begin{align}\label{eq:coupling}
\xi_{nm}^{(0)} &=\mathcal{N}\:\int_{V_{cavity}} (\varepsilon_c-\varepsilon_s) \mathbf{E}_{n}^{0*}(\mathbf{R})\mathbf{E}_{m}^{0}(\mathbf{R}) d\mathbf{R}\,\\
\xi_{nm}^{(slab)} &=\mathcal{N}\:\int_{V_{slab}}(\varepsilon_{slab}^0-\varepsilon_s) \mathbf{E}_{n}^{0*}(\mathbf{R})\mathbf{E}_{m}^{0}(\mathbf{R})\: d\mathbf{R}\,\\
\delta\xi_{nm}^{(slab)} &=\mathcal{N}\:\int_{V_{slab}}(\varepsilon_{slab}-\varepsilon_{slab}^0) \mathbf{E}_{n}^{0*}(\mathbf{R})\mathbf{E}_{m}^{0}(\mathbf{R}) d\mathbf{R}\,\\
 \xi_{nm} &=\xi_{nm}^{(0)}+\xi_{nm}^{(slab)}+\delta\xi_{nm}^{(slab)}\,,
\end{align}
\end{subequations}
where 
\begin{align}
\mathcal{N} = \frac{\omega_m}{2} \left (\int \varepsilon_s |\mathbf{E}_n^{0}(\mathbf{R})|^2 d\mathbf{R} \int \varepsilon_s |\mathbf{E}_m^{0}(\mathbf{R})|^2 d\mathbf{R} \right )^{-1/2}\,.
\end{align}
Using a similar approximation as in case of the nano particle, the change $\delta\xi_{nm}^{(slab)}$ induced is proportional to $\delta\varepsilon_{slab}$ and given by \cite{PRA75p023814,OE13p1515}
\begin{equation}\label{eq:SLAB}
 \delta \xi_{nm}^{(slab)}=\frac{\delta \varepsilon_{slab}}{\varepsilon_{slab}^0-\varepsilon_s}\xi_{nm}^{(slab)}\,.
\end{equation}
Thus a large static coupling $\xi_{nm}^{(slab)}$ is favorable. Again, since the transmission and reflection in our setup depend on the coupling constants, such a variation of the coupling constants can be detected.

\subsection{\label{aver}Reflection and transmission averaged over coupling and scattering phases}

In this Section we study the impact of an averaging over certain phase angles $\phi_{ij}$ on the transmission and reflection spectra. In all cases, we average over the full range of $2\pi$. 
The averaging over the coupling constant phase $\phi_{12}$ could be visualized as an experimental setting in which the refractive index of the medium between cavity 1 and 2 changes between several measurements, e.g., due to changes in the concentration of a fluid filling this region. The averaging over a scattering constant $\phi_{22}$ can be visualized as arising from different particle positions throughout the measurements~\cite{kimble}.

In general, we find that the averaged curves differ considerably from the curves obtained for fixed phases such as $\phi_{ij}=0$.
\begin{figure}
\includegraphics[width=8cm]{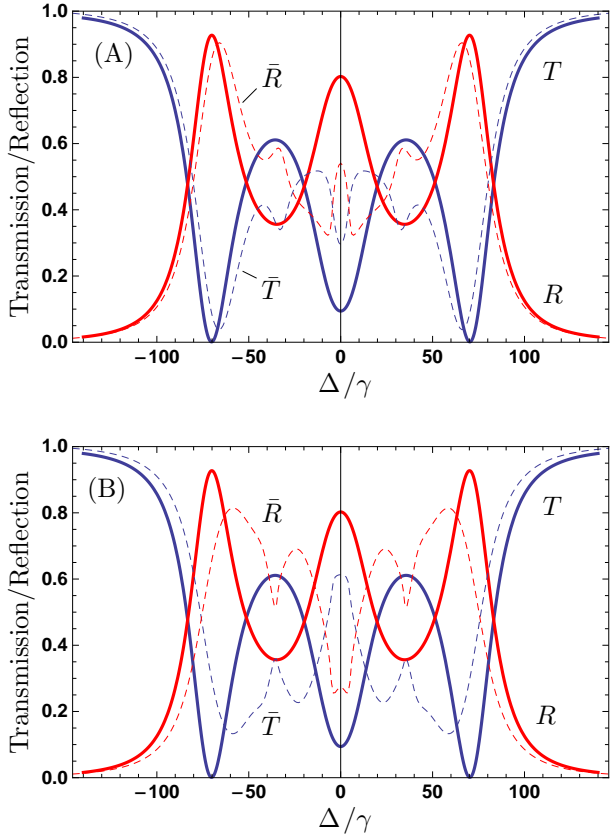}
\caption{\label{av2}(Color online) Transmission $T, \bar{T}$ and reflection $R, \bar{R}$ for parameters as in Fig.~\ref{EW2}. In (A), the solid lines $T,R$ show the case with phases $\phi_{ij}=0$, the dashed lines $\bar{T}, \bar{R}$ show results averaged over $\phi_{22}$. 
In (B),  the solid lines $T,R$ show the case with phases $\phi_{ij}=0$, the dashed lines $\bar{T}, \bar{R}$ show results averaged over $\phi_{12}$. }
\end{figure}
Fig.~\ref{outint}(A) shows that averaging over $\phi_{12}$ or $\phi_{22}$ lead to the same results, as long as resonators 2 and 3 are uncoupled, i.e., $\xi_{23}=0$. This can be understood by noting that in case of $\xi_{23}=0$, all pathways in which $\phi_{12}$ or $\phi_{22}$ lead to a final phase shift of the corresponding amplitude include both an interaction with the coupling $\xi_{12}$ and an interaction with the scattering rate $\xi_{22}$, see Eq.~(\ref{C1}). Therefore either averaging has the same effect.

In contrast, if $\xi_{23}\neq 0$, the spectra averaged over the scattering phase $\phi_{22}$ or over the coupling phase $\phi_{12}$ differ, see Fig.~\ref{av2}. The reason is that now $\phi_{12}$ and $\phi_{22}$ affect different pathways. For example, the pathways in Eqs.~(\ref{D1})-(\ref{pathD}) include a phase contribution of $\phi_{12}$, but not of $\phi_{22}$. Thus in contrast to the case without roundtrip process, the averaging over the two phases leads to different results.

\begin{figure}[t]
\centering
\includegraphics[width=\linewidth]{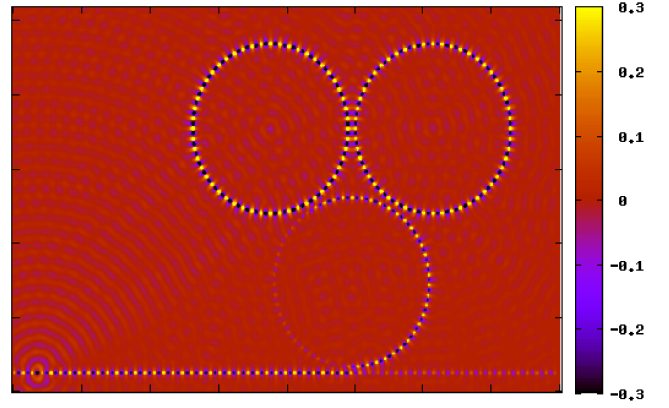}
\caption{\label{fdtd-example}(Color online) Configuration of the field component $H_z$ obtained from the FDTD-simulation for three resonators in loop configuration without particle or slab.  The image shows one snapshot in time. The excitation wavelength is $\lambda = 571.8$~nm. See text for further explanation.}
\end{figure}

 \subsection{Numerical simulations with FDTD}

In this section we numerically verify the suggested applications using a finite-difference time-domain (FDTD) solution of Maxwell's equations on a grid~\cite{FDTDATaflove}. The simulations consider two dimensions for TM modes propagating in $x$ direction (field components $E_x$, $E_y$, $H_z$)  on a Yee-grid with grid size of $30$~nm. The grid boundaries are modeled using Berenger-type perfectly absorbing boundary conditions, and we use a point-like sinusoidally oscillating source in $H_z$. We consider a background with permittivity $\varepsilon_s=1$, and resonators and a waveguide with permittivity $\varepsilon_c=4$. The waveguide resonators have outer radius $3500$~nm and inner radius $3350$~nm. The distance between waveguide and resonator is 120~nm, the distances between the resonators are chosen equal as 200~nm. The waveguide has a width of 150~nm. 

A typical example for three resonators in loop configuration is shown in Fig~\ref{fdtd-example}. The figure shows the field component $H_z$ after the time evolution has reached a stationary state. The pointlike source is in the lower left corner, and is placed in the center of the waveguide which runs along the lower edge of the figure. Since the pointlike source does not exclusively excite waveguide modes, circularly spreading background waves originating from the source can be seen as well. The excitation then runs along the waveguide to the three resonators seen as the circular field arrangements. In this particular example, the two upper cavities contain standing wave excitations, which manifest themselves a modulated total intensity (``blinking'') in the time-dependent dynamics of the field configuration. The right half of the lower cavity exhibits a less pronounced standing wave, whereas the left half is mostly filled by a running wave in clockwise direction. This ``blinking'' can be seen by comparing the figure to a corresponding snapshot slightly later in time. If the time is chosen appropriately, the time evolution of the standing waves is close to a minimum, such that the bright field regions in Fig~\ref{fdtd-example} are almost invisible. In contrast, the running wave parts remain similar. In total, the setup in Fig~\ref{fdtd-example} leads to a weak forward transmission $(T\ll 1)$, which can be seen from the low field excitation downstream of the resonators. The energy is instead mostly reflected, which again is evidenced by standing wave field components in the waveguide between source and resonators. In contrast, the other waveguide parts only carry running wave excitations, as expected. 

\begin{figure}[t]
\centering
\includegraphics[width=\linewidth]{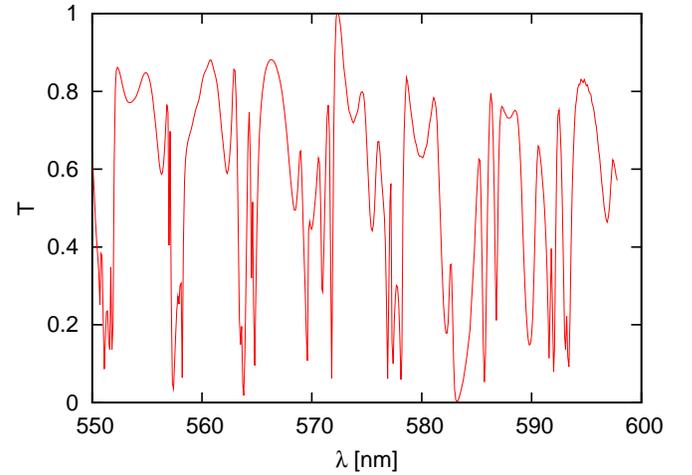}
\caption{\label{fdtd-spectrum}(Color online) Transmission spectrum obtained from FDTD simulations for three resonators in loop configuration without particle or slab.}
\end{figure}

To evaluate the transmission, we sum the time-averaged Poynting vector contributions in a plane transversal to the waveguide to the right of the resonators. The obtained energy flux is normalized to the case without resonators. By varying the frequency of the excitation, a transmission spectrum can be determined. A typical example is shown in Fig.~\ref{fdtd-spectrum}.

\begin{figure}[t]
\centering
\includegraphics[width=\linewidth]{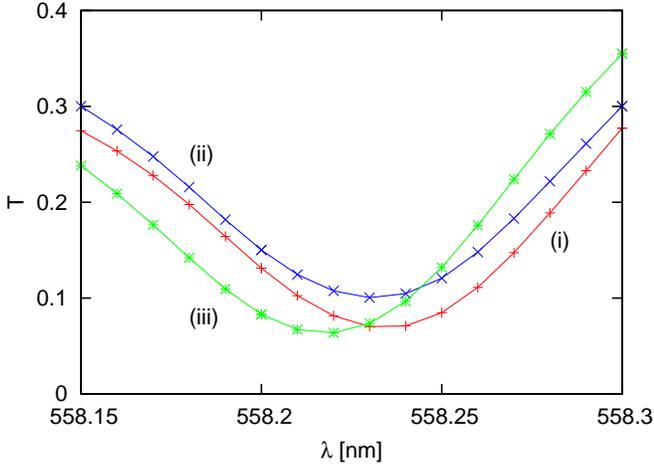}
\caption{\label{fdtd-particle}(Color online) Transmission spectrum obtained from FDTD simulations for three resonators in loop configuration with a particle placed close to the top left resonator. The different curves show particle positions (i) $\theta = 90^\circ$, (ii) $\theta = 95^\circ$, and (iii) $\theta = 180^\circ$. }
\end{figure}

\begin{figure}[t!]
\centering
\includegraphics[width=\linewidth]{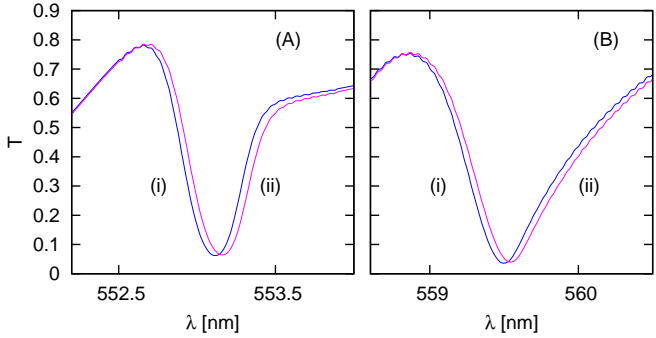}
\caption{\label{fdtd-slab}(Color online) Transmission spectrum obtained from FDTD simulations for three resonators in loop configuration with a slab placed between the two rightmost cavities, see Fig.~\ref{fig:Carton}(B). (A) and (B) show two spectral lines at different wavelengths. (i) corresponds to a slab with permittivity $\epsilon=4.0$, (ii) to a slab with $\epsilon=4.1$. }
\end{figure}

For the applications, we add a particle or a slab to the cavities as indicated in Fig.~\ref{fig:Carton}.  The particle has a radius of $90$~nm, and it is placed at different azimuthal positions with a radial distance of 90~nm to the top left cavity. The transmission spectrum around one spectral peak is shown for different particle positions in Fig~\ref{fdtd-particle}. Other than the azimuthal position, no parameters are changed between the three curves in this figure. It can be seen that changing the position leads to a shift of the resonance line in the transmission spectrum. This shift can be observed over a large range of position values, as indicated by the lines with $90$ degrees and $180$ degrees. On the other hand, a shift can  already be observed for a change of position by $5$ degrees. This has to be compared to the angular range of about $7$ degrees for one wavelength in the azimuthal mode number $52$ realized in this numerical example.

Fig.~\ref{fdtd-slab} shows corresponding results for a slab. The slab has width 60~nm and length 1500~nm and is placed symmetrically between the two rightmost resonators. We consider slabs with permittivity $\varepsilon=4.0$ and $\varepsilon=4.1$. The two subfigures in Fig.~\ref{fdtd-slab} show two resonances in the transmission spectrum. It can be seen that the change in the permittivity from 4.0 to 4.1 shifts the position of a resonance line. It is interesting to note that not the whole spectrum is shifted, but only part of the spectral lines, as can be seen by the parts in  Fig.~\ref{fdtd-slab} which are not affected by the change of the slab.

We thus conclude that the FDTD simulations of the considered loop systems with additional particle or slab serve as a proof or principle for the applications discussed in Sec.~\ref{appl}. While a direct connection to the coupled-mode calculations is not possible since the individual coupling constants and resonator properties realized in the numerical simulations are unknown, both the variation of the particle position and the index of refraction of the slab led to a shift of resonance lines in the transmission spectrum, and therefore should be detectable.

\section{Summary}

We analyzed interference effects in a system consisting of three coupled microcavities arranged in such a way that light can evolve through the resonators in a non-trivial loop roundtrip. The system is probed by a fiber coupled to one of the resonators. The interplay of the different pathways light can take while passing through the resonator array leads to rich structures in the transmission and reflection spectra of the system. In particular, we have focused on a sensitivity of the spectra on the phases of the different scattering and coupling constants. We found that the roundtrip process in which light moves on a circle through all three cavities, enabled by the special arrangement of the resonators, leads to additional pathways which increase the sensitivity of the spectra to the phases. Finally, we discussed two applications for the found phase-sensitivity. First, we studied the determination of the position of a particle placed in the evanescent field of one of the resonators. Second, we analyzed the measurement of the index of refraction of a slab placed between two cavities.  Our results are on the one hand based on quantum mechanical coupled mode theory, which allows to interpret the spectra in detail based on the underlying physical mechanisms. On the other hand, we verified the sensitivity of the spectra in the two discussed simulations using numerical finite-difference time-domain simulations of Maxwell's equations on a grid.


%

\end{document}